\documentclass[aps,prb,twocolumn,superscriptaddress,groupedaddress,footinbib]{revtex4}
\usepackage[dvips]{graphicx}
\usepackage{subfigure}
\usepackage{epstopdf}
\usepackage{color}
\usepackage{ucs}
\usepackage{float}
\usepackage[utf8x]{inputenc}
\begin{document}

\title{Stacking Characteristics of Close Packed Materials}
\author{Christian H. Loach and Graeme J. Ackland}
\affiliation{School of Physics and Astronomy, SUPA, The University of Edinburgh, Edinburgh, EH9 3JZ, United Kingdom}
\date{\today}

\begin{abstract}
It is shown that the enthalpy of any close packed structure for a
given element can be characterised as a linear expansion in a set of
continuous variables $\alpha_n$ which describe the stacking
configuration. This enables us to represent the infinite, discrete set
of stacking sequences within a finite, continuous space of the
expansion parameters $H_n$. These $H_n$ determine the stable structure
and vary continuously in the thermodynamic space of pressure,
temperature or composition.  The continuity of both spaces means that
only transformations between stable structures adjacent in the $H_n$
space are possible, giving the model predictive and well as
descriptive ability.  We calculate the $H_n$ using density functional
theory and interatomic potentials for a range of materials.  Some
striking results are found: e.g. the Lennard-Jones potential model has
11 possible stable structures and over 50 phase transitions as a
function of cutoff range. The very different phase diagrams of Sc, Tl,
Y and the lanthanides are undrstood within a single theory.  We find
that the widely-reported 9R-fcc transition is not allowed in
equilibrium thermodynamics, and in cases where it has been reported in
experiments (Li, Na), we show that DFT theory is also unable to
predict it.

\end{abstract}
\maketitle

In 1611, Kepler suggested that stackings of
triangular layers was the most efficient way to pack hard spheres
\cite{kepler}.  This conjecture was only recently
proved\cite{hales}.

Many elements
crystallise in close-packed crystal structures, but the concept of 
``close-packed'' is not part of crystallographic categorization. 
This is because there are an infinite number
of stacking arrangements with equal packing density, spanning a wide range of
space group symmetries.  Most observed structures have short
repeat sequences such as face-centered cubic (fcc) or hexagonal close
packed (hcp), but there is no general theory to explain why these
should have the lowest energy.

Predicting the stable crystal structure for a material is a
longstanding challenge in condensed matter physics.  One underlying
reason is that crystal structures are defined by discrete symmetry
groups and integer numbers of atoms per unit cell.  Aside from the
atomic positions themselves, there are no continuous variables which
cover the entire space of possibilities, thus we are searching for a
minimum in a discontinuous space.

Among close-packed structures, only fcc has close packing enforced by
symmetry.  For all other stackings, there is an ``ideal'' ratio
between interlayer spacing and interatomic separation  ($c/a=\sqrt{2/3}$)
which gives close-packing. Generally, materials adopting structures within a
few percent of ``ideal'' are regarded as close-packed.  

Stacking sequences are typically defined as a series of layers labelled
$\mathtt{A}$, $\mathtt{B}$, and
$\mathtt{C}$ with atoms positioned at $0 \mathbf{a} + 0 \mathbf{b}$;
$\frac{1}{3}\mathbf{a} + \frac{1}{3}\mathbf{b}$; and
$\frac{2}{3}\mathbf{a} + \frac{2}{3}\mathbf{b}$ respectively, where
$\mathbf{a}$ and $\mathbf{b}$ are the in-plane lattice vectors.
This $\mathtt{ABC}$ notation is not unique: a
more compact notation \cite{Christian1970} uses $\mathtt{h}$ for
layers with identical neighbours ($\mathtt{ABA}$), $\mathtt{f}$ for those with
different ($\mathtt{ABC}$).  
For examples see table \ref{table:basics}.

The most widely-used model for atomistic modelling is the
Lennard-Jones potential, which describes the van der Waals bonding of
inert gases.  It has hcp as the most stable structure at low temperature, 
transforming to fcc at high  temperature\cite{jackson2002lattice}.  More
sophisticated modelling of electronic structure using density
functional theory can be applied across the periodic table, and gives
quantitative agreement with experiment\cite{Jain2013} although it is
impossible to check all possible stacking sequences.

In this paper we show that the energies of the infinity of stacking
sequences can be represented by a convergent series, and that phase
boundaries between some pairs of crystal structures cannot occur.  We
demonstrate the extraordinary complexity of the Lennard-Jones phase
diagram.  We show that deviations from ``ideal'' c/a ratios are
correlated with stability.  We also investigate the role of pressure
and uncover some deep-seated inadequacies in interatomic potentials.

\begin{figure}[htb]
\begin{center}
\includegraphics[width=0.41\linewidth]{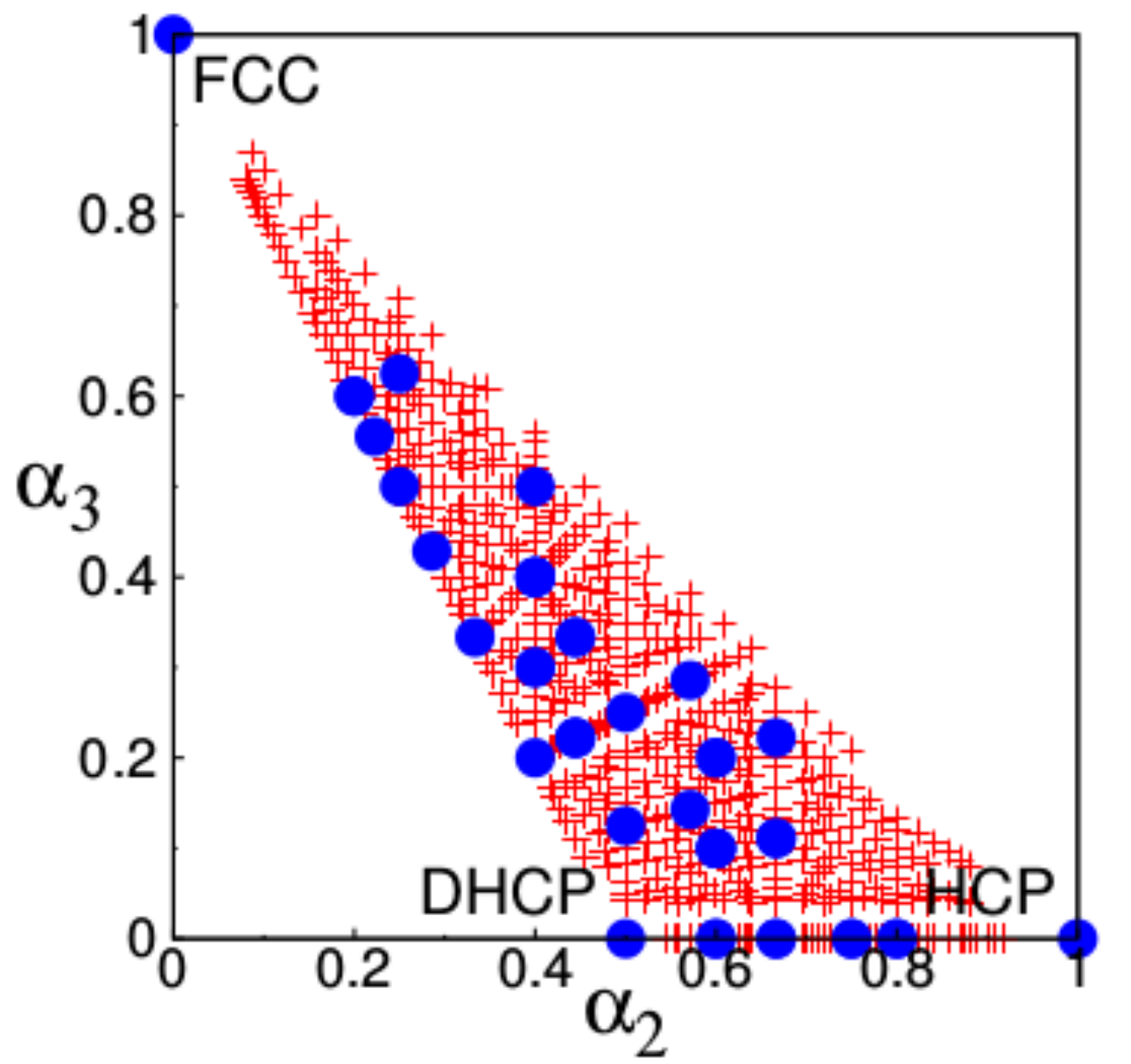}
\includegraphics[width=0.57\linewidth]{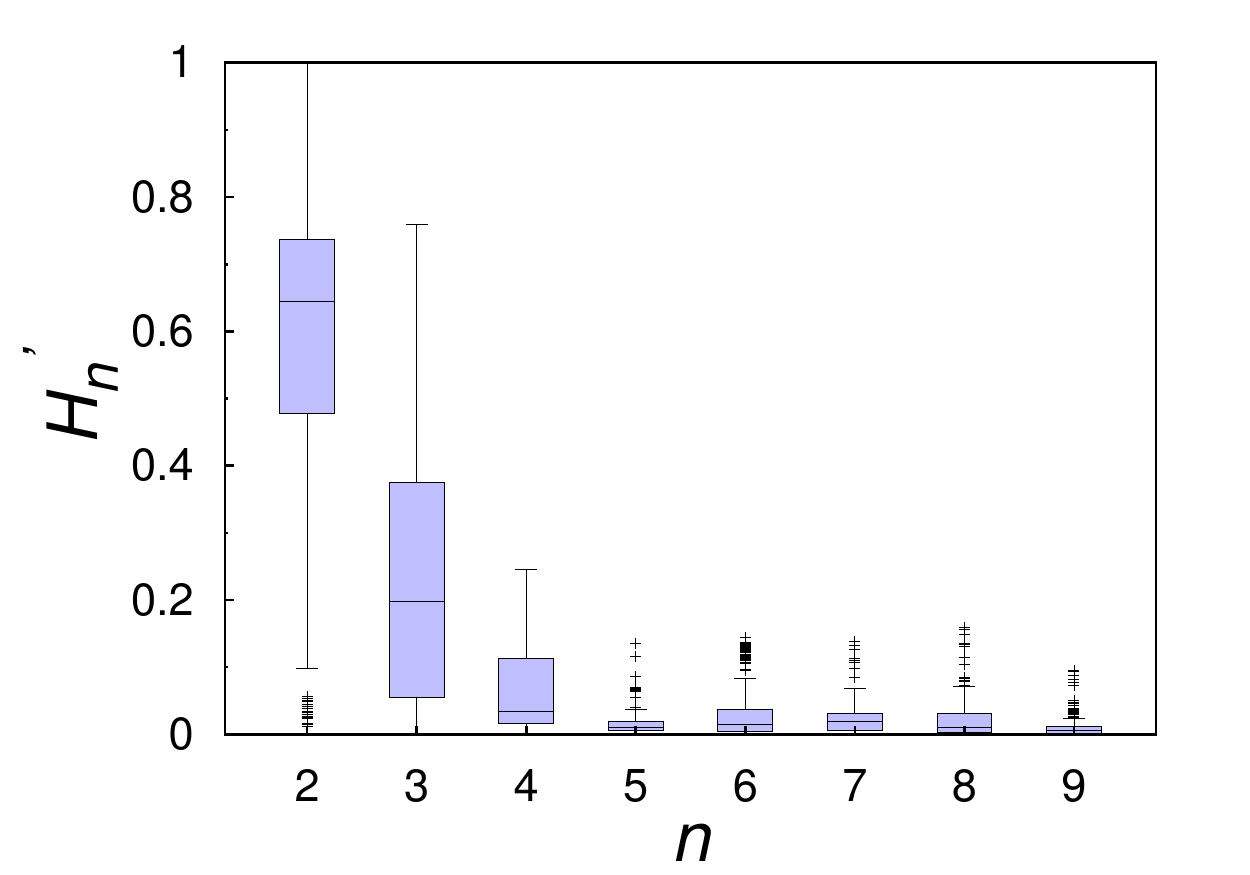}
\caption{(left) Physically realizable stackings projected  onto the
  $\alpha_2$-$\alpha_3$ plane. Configurations for up to 25 atomic
  layer repeats are shown in red. Blue points indicate the 43 structures
 used in our
  calculations. (right) Box plots of normalised enthalpy $H^{\prime}_n$ vs $n$
  showing the rapid convergence of
  Eq. \ref{equation:energy}.  Data is taken from DFT
  calculations across all elements and pressures.  The
  structure-independent $H_0$ are omitted.   \newline 
 Specifically
$H^{\prime}_{n} = \frac{\mid H_{n} \mid}{\sum_{i=2}^9 \mid H_{i}
    \mid}.$}
\label{fig:phasespace}
\end{center}
\end{figure}

To define the stacking sequence with periodicity M, we introduce a set of
parameters $\alpha_n$

\begin{equation}  
\alpha_n =   \sum^M_{i=1} \frac{\delta_{i,i+n}}{M} 
\end{equation}

where 
$\delta_{i,i+n}$ is 1 when the $i$ and $i+n$ layers have the same ABC
symbol, and 0 otherwise.  Physically $\alpha_n$ can be thought of as
``The fraction of the atomic positions ${\bf R_i}$ for which there is
another atom at ${\bf R}_i + n{\bf c}$'', where $c$ is the interlayer
separation. As $M\rightarrow\infty$, or for an arbitrary density of
stacking faults, the $\alpha$s become continuous variables,

The set of $\alpha$'s up to $\alpha_M$
univocally describes any possible stacking with an $M-$fold or fewer
periodicity.  All translationally, rotationally or reflectionally
equivalent stackings have the same unique set of $\alpha_n$,
unlike the ABC and hf notations which have considerable redundancy.  
Trivially, $\alpha_0=1$ and $\alpha_1=0$ for all close-packed structures.
Only certain ranges of $\alpha_n$s correspond to physically-realizable structures (see Fig.\ref{fig:phasespace}).

\begin{table}[htb]
\begin{center}\vspace{0.2cm}
\begin{tabular}{llllll}
\toprule 
Name & \textbf{ABC} & \textbf{hf}  & Minimal
 &\textbf{$\alpha_2$} & \textbf{$\alpha_3$} \\ \hline\vrule
height 12pt width 0pt 
hcp & $\mathtt{AB}$ & $\mathtt{hh}$ &$\mathtt{h}$ & 1
 & 0 \\ fcc &$\mathtt{ABC}$ & $\mathtt{fff}$ &$\mathtt{f}$ & 0 & 1
\\ fcc&  $\mathtt{ACB}$ & $\mathtt{fff}$ &$\mathtt{f}$ & 0 & 1
\\ dhcp & $\mathtt{ABCB}$ & $\mathtt{hfhf}$ &$\mathtt{hf}$ & 1/2 & 0
\\ & $\mathtt{ABCAB}$ & $\mathtt{hfffh}$ &$\mathtt{hfffh}$ & 2/5 & 2/5
\\ 9R & $\mathtt{ABACACBCB}$ & $\mathtt{hhfhhfhhf}$ &$\mathtt{hhf}$ & 2/3 & 0
\\ \hline
\end{tabular}
\caption{Representation of various structures in terms of basal
  stacking in the different notations.  Note that 
 $\mathtt{ABC}$ and
  $\mathtt{ACB}$ represent the same structure, fcc and that structures
  are not uniquely defined by $\alpha_2$, $\alpha_3$.}
\label{table:basics}
\end{center}
\end{table}

Utilizing the CASTEP simulation package \cite{Clark2005},
well-converged energies for various stackings were determined in the
framework of density functional theory using the PBE
exchange-correlation functional\cite{PBE} for a selection of elements
known to adopt close packed structures at a range of pressures.  In
addition to the DFT calculations, we calculate energies of the same
structure set using a number of  interatomic potentials,
both pairwise and many-body, which were fitted to represent the same
materials.  Our structure set consists of all 43 possible stacking
sequences for up to 10 atomic layer repeats in the $\mathtt{ABC}$
notation (c/f Table \ref{table:basics}) 
excluding redundant strings (i.e. those with identical $\alpha_n$). 
Calculations are performed starting from
hexagonal style unit cells with cell angles 90\textdegree,
90\textdegree, 60\textdegree; Internal coordinates and lattice
parameters were fully relaxed, and double-checked to ensure that each
structure remained in its initial metastable state, with each atom in
the structure retaining 12-fold coordination and undergoing only small
distortion from close-packing.

Each material is characterized by parameters $H_n$ 
which are obtained by a least squares fit to the 43 calculated enthalpies 
assuming a linear dependence on $\alpha_n$, 
\begin{equation}
H = H_{0} + \sum_{n=2} H_n \alpha_n
\label{equation:energy}
\end{equation}
Every material is therefore represented as a point in an N-dimensional
H$_n$-space, and every point in the H$_n$-space has an associated
most-stable stacking structure calculated by minimizing
Eq.\ref{equation:energy} with respect to $\alpha_n$. 
e.g. consider the
summation in Eq.\ref{equation:energy} up to only $n=3$, 
the enthalpy varies linearly with $\alpha_2$ and $\alpha_3$, and
it follows that the most stable structure must
be located at a corner of the triangle of physically-possible states shown in
figure \ref{fig:phasespace}(a), allowing only fcc, hcp, or dhcp. More
complex structures may be stable if considering $H_4$ and higher
terms.

\begin{figure}
\begin{center}
(a)\includegraphics[width=1.0\linewidth,height=0.625\linewidth]{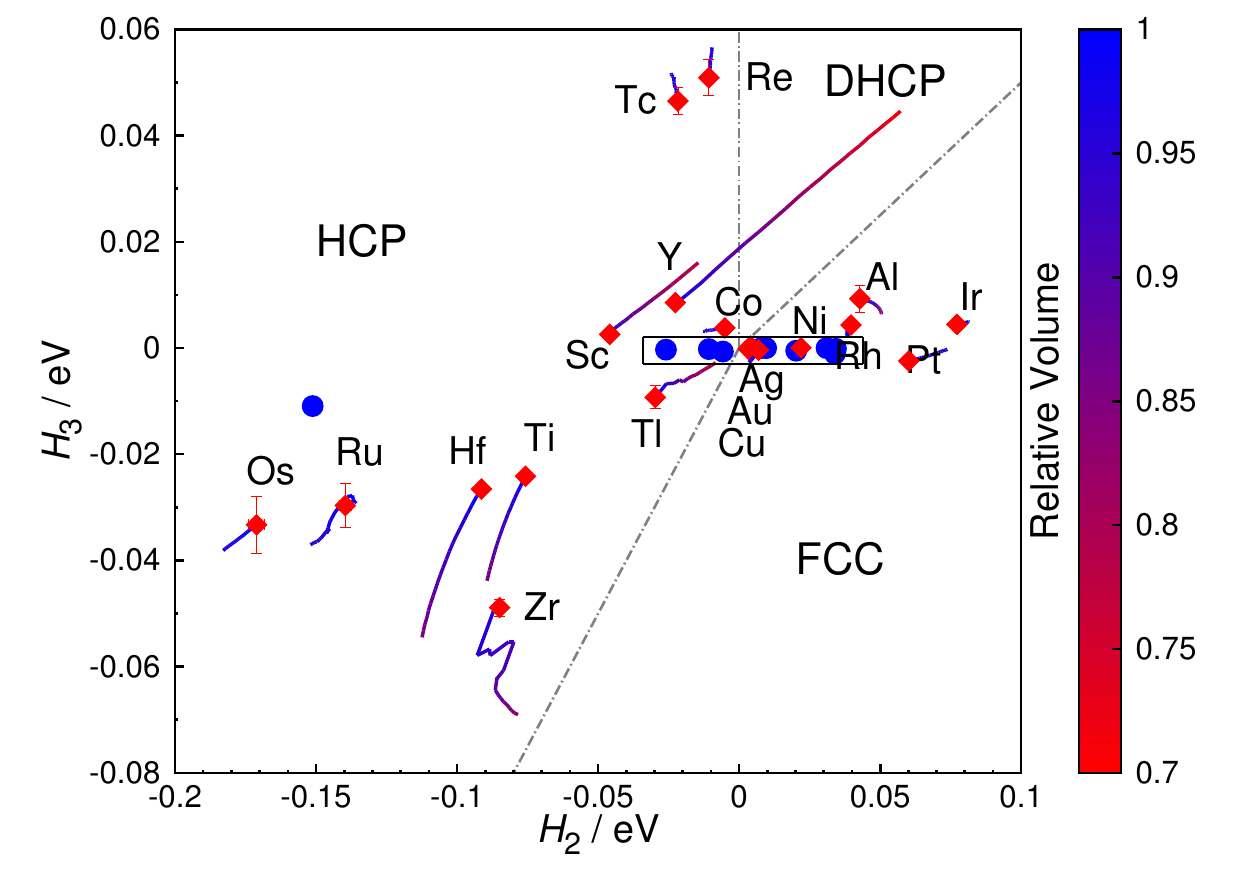}
(b)\includegraphics[width=1.0\linewidth,height=0.625\linewidth]{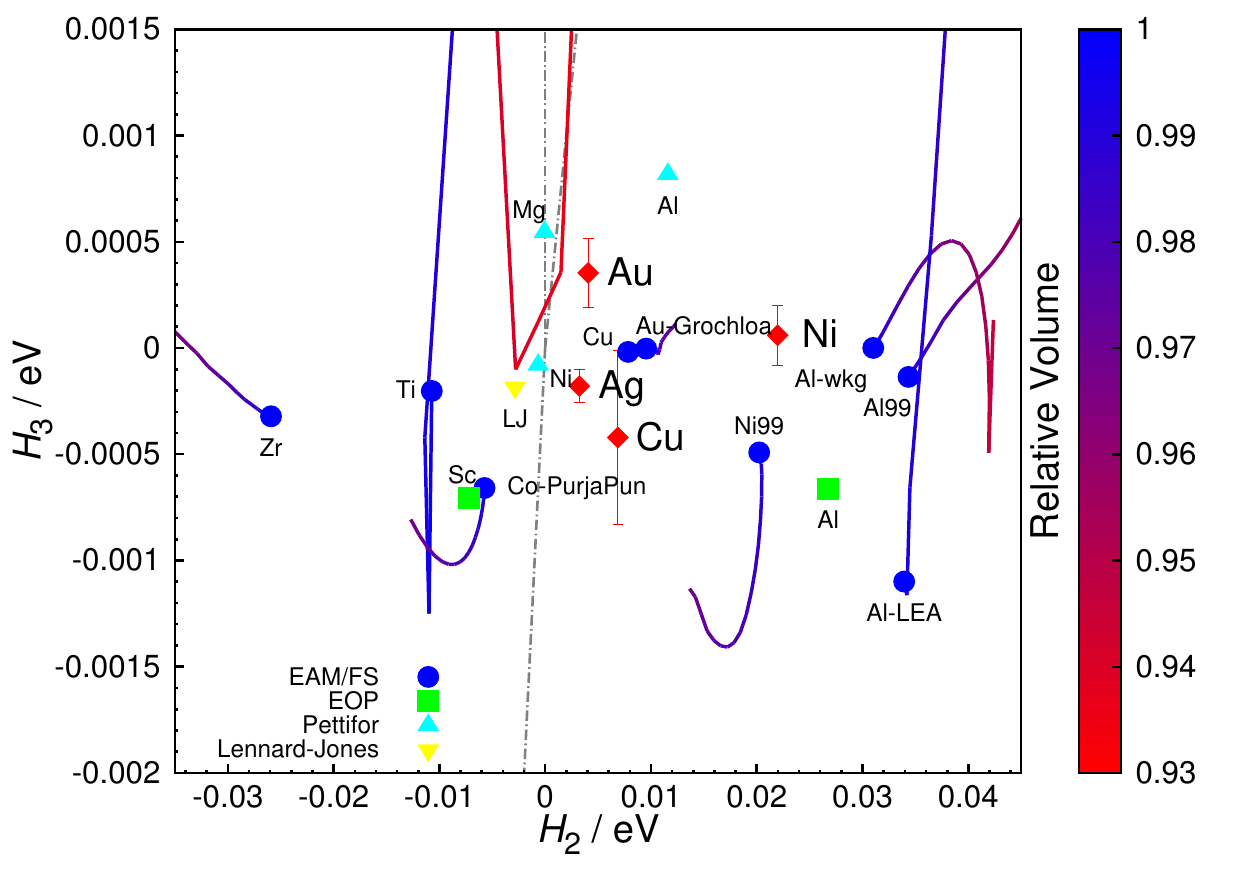}
\caption{(a) Figure showing close packed materials plotted against
  their $(H_2, H_3)$. Lines show the movement under pressure according
  to DFT calculations. Blue dots show the position of interatomic
  potentials at equilibrium volume.  The outlying interatomic
  potential is Fortini's Ru EAM potential \cite{Ru}. The regions of
  fcc, hcp and dhcp stability are shown, boundaries calculated for the
  slice where $H_4$ and higher terms are zero.   
(b) Expanded view of the position
  of interatomic potentials in the region of $H_2$-$H_3$ space bound
  by the rectangle in (a). The lines again show the effects of
  compression.  }
\label{fig:phasegraph}
\end{center}
\end{figure}

The $H_2$ and $H_3$ values for a range of materials and pressures are shown in
Fig. \ref{fig:phasegraph}(a).  The residuals in the fit to DFT data
are of order tenths of meV per atom, about 1\% of the enthalpy
differences between structures.
For Eq.\ref{equation:energy} to be useful it must be rapidly convergent, and in
Fig \ref{fig:phasespace}(b) we show that the terms do indeed decay
rapidly with $n$.   Typically, the $H_{2}$ and $H_{3}$ contributions are dominant.

\begin{center}
\begin{figure}
\includegraphics[width=1.0\linewidth]{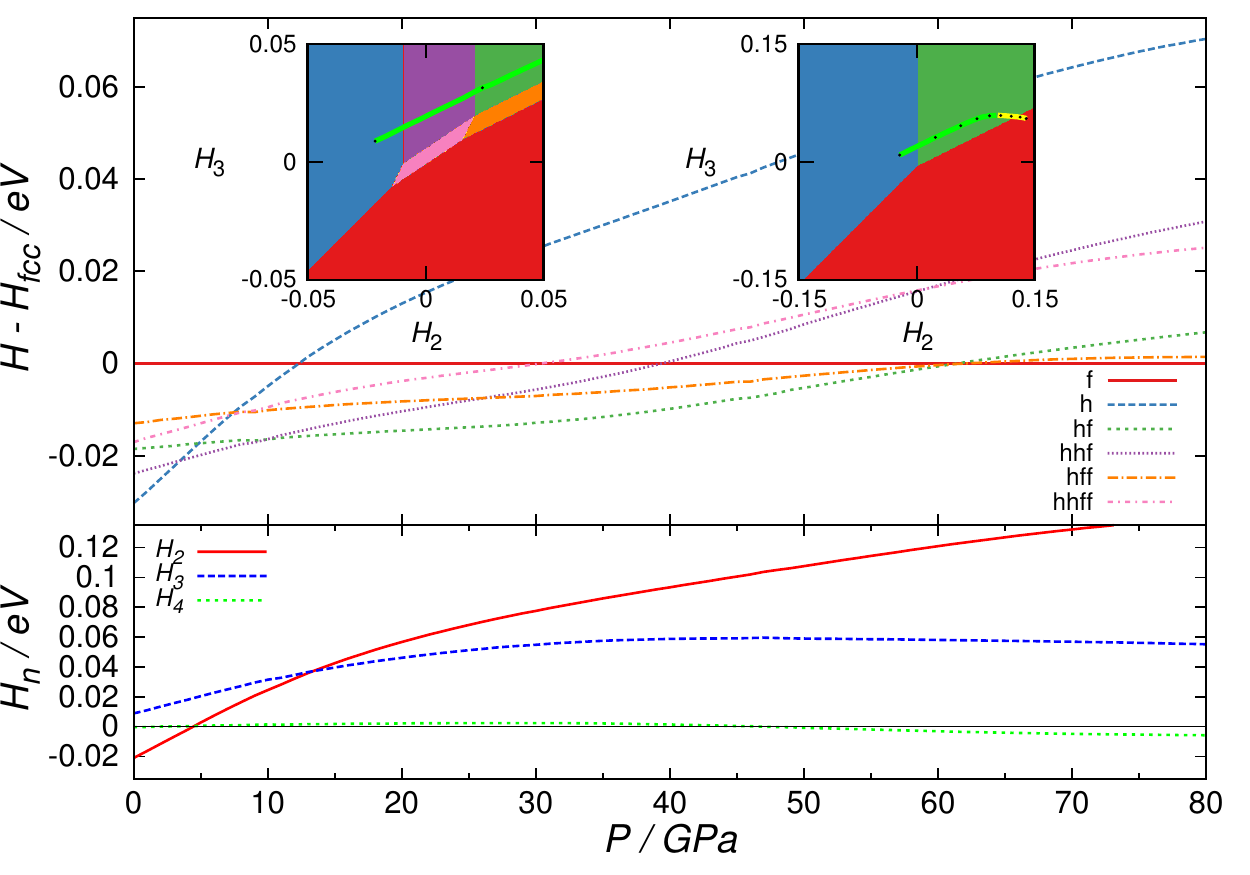}
\caption{(top) DFT calculated enthalpies for phases of Yttrium with
  pressure. (bottom) Fitted $H_n$ values with pressure.  The insets
  are colored to show the stable phase for given ($H_2$, $H_3$) using
  the same color scheme; when $H_4$ is positive (left), all six phases
  appear, for negative $H_4$ (right) only fcc, hcp and dhcp are
  possible.  The line shows changing values of ($H_2$, $H_3$) 
  with pressure.  Because $H_4$ for Y is also
  pressure-dependent, this is a projection onto the plane of constant
  $H_4$ which it intersects: the line is colored green when the $H_4>0$
  and yellow when  $H_4<0$ to show
  that it passes through the wedge of $hhf$ stability, but not $hff$.  Small
  dots indicate 10 GPa intervals.}
\label{fig:Y}
\end{figure}
\end{center}

The key to the usefulness of this result is that we have transformed
the discrete representation ($\mathtt{ABC}$ or $\mathtt{hf}$) of the
crystal structure to a continuous space one ($\alpha_n$).  This
enables us to anticipate phase transitions arising from continuously
changing thermodynamic variables such as temperature, pressure or
composition.  To do this, consider the N-dimensional $H_n$ space.  Any
stacking will have some region of stability if $N$ is large
enough\cite{fisher1980infinitely}.  Geometrically, these regions are
hyperpyramids which meet at the origin where enthalpy is independent
of stacking.  If we change the pressure continuously, the $H_i$ also
change continuously, tracing a path through the $H_n$-space which can
be evaluated based on DFT calculations at different pressures a given
material.  When this path crosses from the stability region of one
phase to another, this corresponds to a phase transition.  A dramatic
physical consequence is that transformations between phases whose
stability regions are non-adjacent in $H_n$ space (Fig \ref{fig:Y}),
such as fcc and 9R, are not thermodynamically possible {\it in any
  system} for which the $H_n$ representation converges.  If the $H_n$
are fitted to free energy calculations, temperature-driven transitions
can also be intimated.

There are similarities with the long-ranged 1D Ising model\cite{bak1980ising,fisher1980infinitely,yeomans1988theory}, in which possible stackings 
(here $h$ and $f$) are represented by spins
\cite{cheng1988inter,cheng1987confirmation,denteneer1987stacking,plumer1988ising,vitos2006evidence}.  In that case $H_2$ maps to
the field, while the Ising interaction terms are linear combinations
of our $H_i$. The Ising representation turns out to be less
useful because it converges slowly.  To understand why, consider the
strings ABACB and ABABC, which give .hff. and .hhf. for the Ising
representation.  In the first case the next neighbour hf interaction
is between unlike (BC) layers, in the second between like (BB) layers.
In the physical system, the set of separations between atoms in B-C is
different from B-B, and the associated enthalpy differences are well 
represented by $H_i$.  In the Ising
picture, this difference emerges from correlations between longer
range interactions, which have an unintuitive mathematical origin.

For a given material, the $H_n$ vary continuously with pressure, temperature or,
for alloys, with composition. Fig.2(a) shows trajectories projected into ($H_2$,
$H_3$) space for pressures up to 20 GPa. 
The
clustering of elements' $H_2$ and $H_3$ values and the similarities of
their pressure dependence corresponds to periodic table groupings,
indicating an electronic origin of the observed properties. 

Many further inferences can be drawn from thr $H_n$ space, for example
Group 11 metals lie close to the origin, and low values of $H_n$
suggest changes in $\alpha$ are not energetically costly. As a
consequence, stacking faults (incremental change in $\alpha_n$) have
low energy, meaning that dislocations can glide easily and Group 11
materials are soft and malleable.

The set of $\alpha_{n}$ 
describe the relationship between close-packed layers, so
non-close-packed phases such as bcc or the $\omega$ phase of titanium
are not accounted for.

Yttrium is a particularly interesting case. 
Projection of its pressure trajectory 
onto the ($H_2$,
$H_3$) plane 
moves it from hcp stability into the dhcp phase
(see Fig.\ref{fig:phasegraph}).
Experimentally\cite{vohra1981structural,Y}, yttrium does this via and
intermediate Sm-type phase, also called 9R, which consists of 9
layers: $\mathtt{ABACACBCB}$ and can be described in the hf notation
as $\mathtt{hhf}$ (Table \ref{table:basics}). 
However hcp, 9R, and dhcp all have $\alpha_3$ values of
zero, and are hence degenerate in situations where $H_2 =0$. 
Consequently 9R lies on the boundary of the hcp phase
with the dhcp phase in figure \ref{fig:phasespace}.  Once $n=4$ terms
are included in equation (\ref{equation:energy}), there is a wedge of 9R
stability for $H_4>0$. This must be traversed as an intermediate
phase between hcp and dhcp, as observed. 

Qualitatively, we find that yttrium transforms from hcp to  9R
at 4 GPa, then to dhcp at around 10 GPa
(Fig.\ref{fig:Y}). These numbers agree with other DFT calculations\cite{chen2011p,chen2012predicted} but are lower
than observed experimental pressures, which might be due
to hysteresis since the experiments were done with increasing pressure
only. 

Scandium and thallium appear to behave similarly to yttrium (see
Supplemental materials), but Sc is known to transform to a complex
non-close-packed structure at a lower pressure than where its trajectory would 
cross the hcp-dhcp boundary
in figure 2(a).  The trajectory for thallium goes towards the
transition line with pressure, but $H_4<0$ so it 
passes below the origin and hcp-fcc is the only observed transition.

The 9R and fcc structures are not adjacent in Fig
\ref{fig:phasespace}.  Therefore, no thermodynamic
phase boundary can exist between 9R and fcc.  This
prohibition of pressure-driven transitions {\it in any
system} is curious because such transitions have been reported in
lithium and sodium.  However, Li 9R was very recently proved not to be
stable\cite{ackland2017quantum}, and we find both Li and Na
to be more stable in fcc than 9R at all pressures.  By
contrast, the 9R phase is adjacent to hcp and dhcp,
(Fig. \ref{fig:phasespace}), so its presence in the samarium phase
diagram is expected. Interestingly, the lanthanide sequence of structures
dhcp/9R/hcp/fcc\cite{gschneidner1968concerning,holzapfel1995structural} is also consistent with
the model.

\begin{figure}
\begin{center}
\includegraphics[width=1.0\linewidth,height=0.625\linewidth]{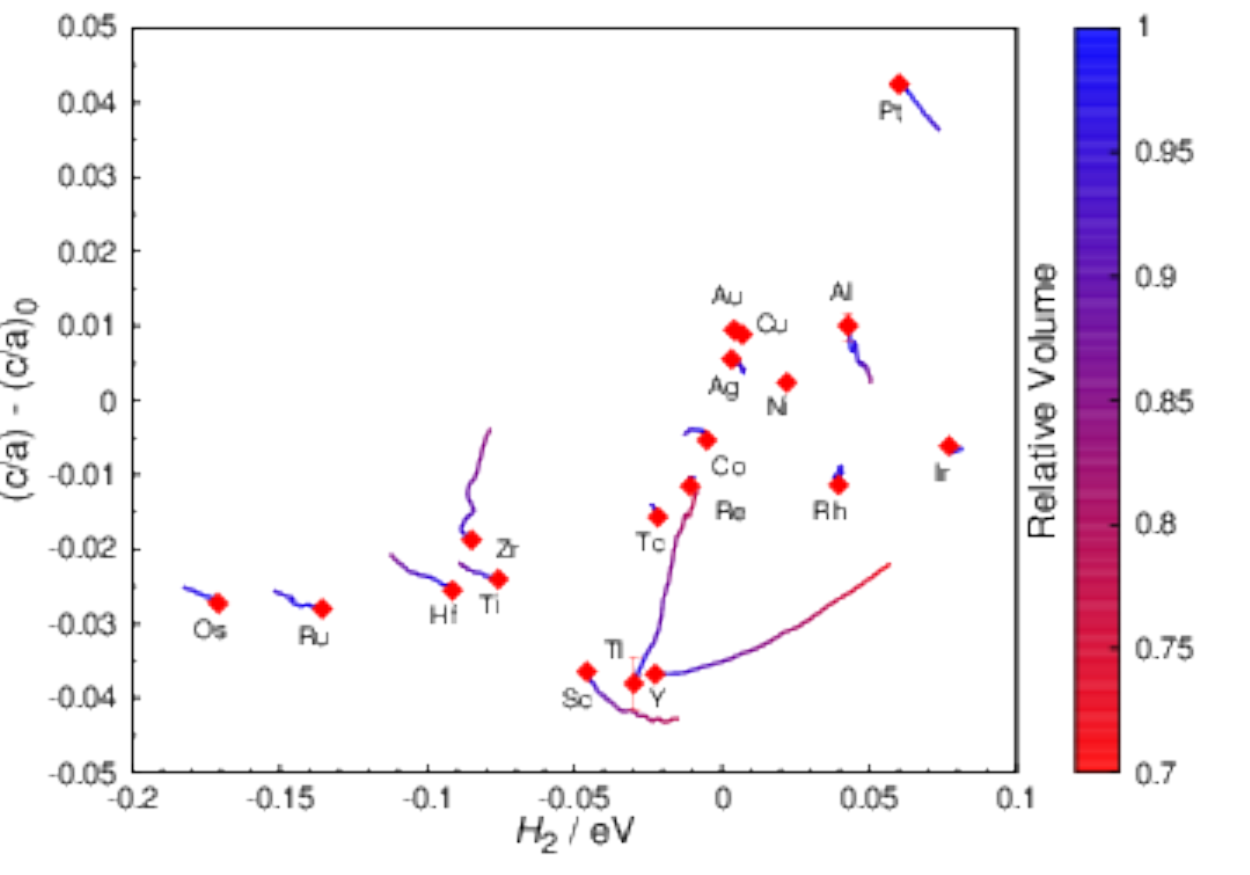}
\caption{Correlation between the stability of hcp over fcc ($H_2$) and the 
divergence from the
ideal close-packed ratio of $(c/a)_0 = \sqrt{\frac{2}{3}}$. The
effect of pressure up to 20 GPa is again shown as paths coloured to
correspond to the relative volume.}
\label{fig:graph2}
\end{center}
\end{figure}

Figure \ref{fig:graph2} shows that the $c/a$ ratio is strongly 
correlated with a material's preference
for the hcp or fcc phase ($H_2$).  Typically, $hcp$ materials
have $c/a<\sqrt{2/3}$, whereas
metastable structures of fcc materials
have larger than ideal $c/a$.  
Curiously, the primary effect of pressure
is to move $c/a$ towards ideal, irrespective of the change in
$H_2$ (Sc being an exception).

The $H_2$ and $H_3$ values for a selection of
interatomic potentials are displayed alongside the
first principles data (Fig \ref{fig:phasegraph}).  We used the Lennard Jones potential, a
set of embedded atom and Finnis-Sinclair potentials
\cite{Au,Ti92,Ru,Al-Ni-99,Al-wkg,Al-lea,Cu,Co}, the Empirical
  Oscillating Potential\cite{Mihalkovic2011}, and Pettifor's three
term oscillating potential for Al, Na, and
Mg\cite{pettifor1984analytic,pettifor1995bonding} as
implemented
in the LAMMPS code\cite{Plimpton1995}.
Remarkably, these potentials almost all fall
into a narrow region of Fig.\ref{fig:phasegraph}(a), shown expanded in
Fig.\ref{fig:phasegraph}(b), the spread on $H_3$ being some two
orders of magnitude smaller than for the DFT calculations.

\begin{figure}[htb]
\begin{center}
\includegraphics[width=1.0\linewidth]{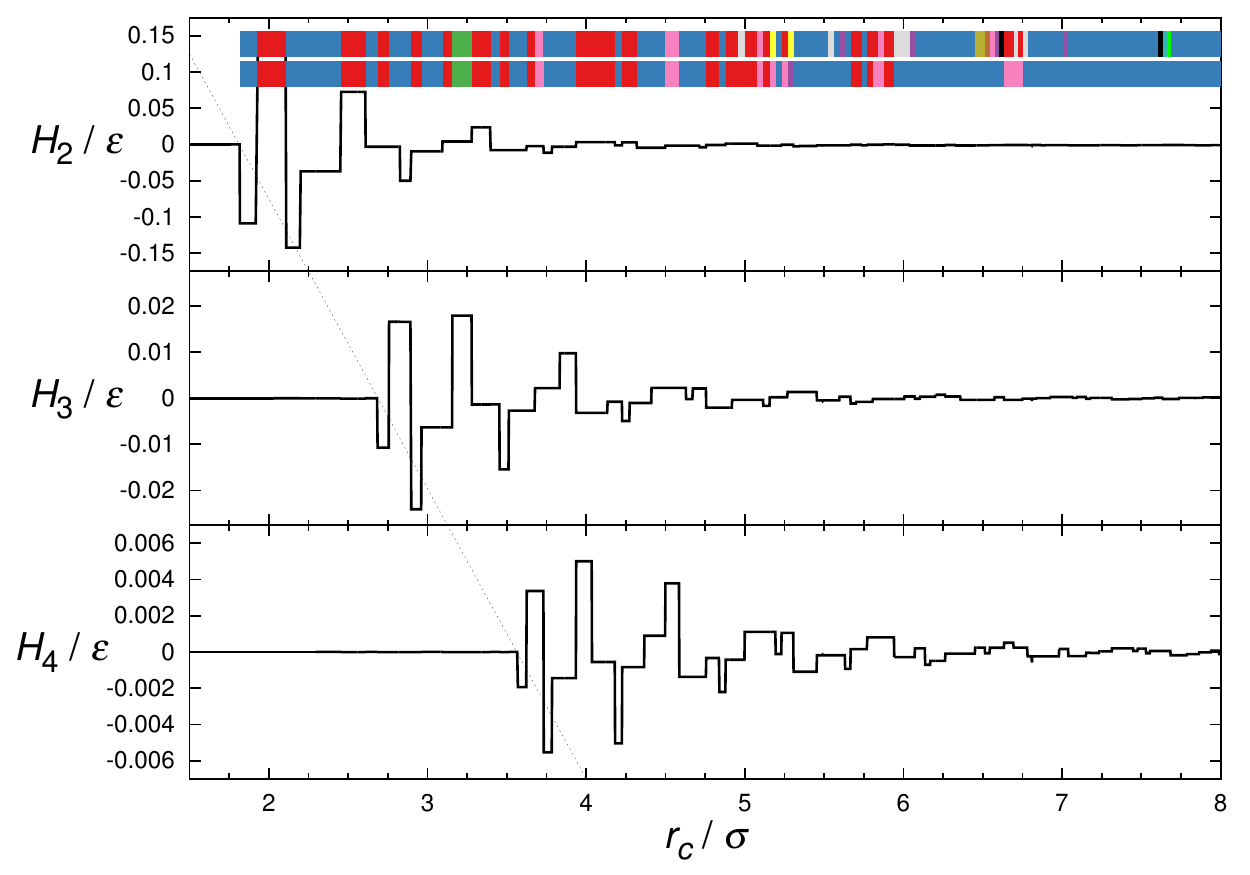}
\caption{Zero-pressure $H_2$, $H_3$, and $H_4$  for the Lennard-Jones
potential as a function of the interaction range. 
The diagonal dotted line demonstrates the regular introduction of new $H_i$ series at intervals of the interplanar spacing. The upper of the two ribbons at the top of the graph shows the minimum enthalpy structure at each value of the cutoff, the lower shows the minimum enthalpy structure predicted by equation \ref{equation:energy}
using the $H_n$ values  up to $n = 4$, . The different colors represent different structures described using the hf notation as follows; Red: $\mathtt{f}$, Blue: $\mathtt{h}$, Green: $\mathtt{hf}$, Purple: $\mathtt{hhf}$, Yellow: $\mathtt{hhhf}$, Pink: $\mathtt{hhff}$, White: $\mathtt{hhfff}$, Olive: $\mathtt{hhhhhf}$, Lime: $\mathtt{hhhhff}$, Cyan: $\mathtt{hhhhhhf}$, Brown: $\mathtt{hhhhfff}$, Black: $\mathtt{hhffhhf}$.}
\label{fig:ljcut}
\end{center}
\end{figure}

This weak dependence of  enthalpy on stacking sequence implies
low basal-plane stacking faults, which leads to
systematic erroneously low barriers to basal slip.  Furthermore, the phase
stability is highly sensitive to pressure and to the details of the empirical
potentials. 

We find truly remarkable results for the Lennard Jones 6-12 forcefield
(Fig \ref{fig:ljcut}).  This most widely-used of potentials is in
practice invariably applied with truncation\cite{Plimpton1995}, at some 
range ${r_{cut}}$.  i.e.
\begin{equation}
\phi(r) = 4\epsilon \bigg[\bigg(\frac{\sigma}{r}\bigg)^{12} - \bigg(\frac{\sigma}{r}\bigg)^{6}\bigg]H(r_{cut}-r)
\end{equation}   
with H the Heaviside function and $\epsilon$ and $\sigma$
defining length and energy units. As ${r_{cut}\rightarrow\infty}$, $H_2$
converges to a value of around $-0.0009 \epsilon$, which accounts for
most of the difference in energy between the fcc and hcp phases, while
$H_3$ converges to a value two orders of magnitude smaller, indicating
a stable hcp ground state. The dependence of the $H_n$ values on $r_{cut}$  
is erratic; discontinuities occur as new
coordination shells come within range, with even $H_2$ changing five times.  This means that a large number of minimum
enthalpy phases are observed as a function of the cutoff, as indicated
in figure \ref{fig:ljcut}.  Calculation using an alternative
truncation with the energy and force shifted to remove the
discontinuities at the cutoff distance, is better behaved, but still
undergoes five transformations with increasing cutoff, with regions of
fcc, hcp and dhcp phases (see Supplemental Materials).

The interatomic potentials exhibit more pressure induced phase
transitions than the DFT calculations.  We propose that this is
because they have a fixed characteristic lengthscale associated with
the zero pressure fitting data.  In reality, the
characteristic length for metallic interactions 
might  be the Fermi wavelength, which reduces with pressure. 
The long ranged oscillations 
of Pettifor potentials scale with
the Fermi vector, meaning that the position of shells of neighbouring
atoms is unchanged relative to the maxima and minima of the
potential\cite{pettifor1995bonding}.  Consequently,
Pettifor potentials show fewer
pressure-induced transitions than other models. 

In summary, we showed that different stackings of
monatomic close packed metals can be uniquely described by 
a set of structure-specific continuous variables $\alpha_n$, and 
that an enthalpy expansion in these quantities leads to a
multidimensional $H_n$ space containing regions of stability for all stackings.
The material-specific fitted expansion coefficients $H_n$ converge
quickly with $n$, and allow the stablest structure to be
determined. Changes in $H_n$ with pressure allow us to identify phase
transformations.

Using the model, we predict that a boundary between fcc and 9R
($\alpha-$Sm-type) phases cannot exist in any phase diagram, requiring
a reassessment of stability of the reported 9R in Na and Li, but not
in the Sm prototype. We reproduce and interpret the phase
transformation sequence in Y, Sc, and Tl.  We identify excess
polytypism as problematic for simple interatomic potentials in general, and  
demonstrate an unprecedented amount of polytypism in the Lennard-Jones system.

\begin{acknowledgements}
GJA acknowledges support from a Royal Society Wolfson award, and the ERC grant Hecate. CHL thanks
EPSRC CM-CDT grant EP/L015110/1 for a studentship, computing resource
was provided by EPSRC via UKCP grant EP/P022790/1.  Data from the calculation is available in Edinburgh datastore 
\end{acknowledgements}

\bibliographystyle{apsrev} 
\bibliography{test} 
\end{document}